\documentclass[sigconf]{acmart}

\AtBeginDocument{%
  \providecommand\BibTeX{{%
    \normalfont B\kern-0.5em{\scshape i\kern-0.25em b}\kern-0.8em\TeX}}}

\setcopyright{iw3c2w3}

\usepackage{xcolor}
\usepackage{multirow}
\usepackage{array}
\usepackage{rotating}
\usepackage{graphicx}
\usepackage{booktabs}

\usepackage[T1]{fontenc}
\usepackage{lmodern}

\iffalse
\newcommand{\raNote}[1]{{\color{blue}[{{#1}}]}}
\newcommand{\rmNote}[1]{{\color{green}[{{#1}}]}}
\newcommand{\pmNote}[1]{{\color{orange}[{{#1}}]}}

\else
\newcommand{\raNote}[1]{}
\newcommand{\rmNote}[1]{}
\newcommand{\pmNote}[1]{}

\fi

\begin{document}

\title{Reviews in motion: a large scale, longitudinal study of review recommendations on Yelp}

\author{Ryan Amos}
\affiliation{%
  \institution{Princeton University}
  \city{Princeton}
  \state{NJ}
  \country{USA}}
\email{rbamos@cs.princeton.edu}

\author{Roland Maio}
\affiliation{%
  \institution{Columbia University}
  \city{New York}
  \state{NY}
  \country{USA}}
\email{rjm2212@columbia.edu}

\author{Prateek Mittal}
\affiliation{%
  \institution{Princeton University}
  \city{Princeton}
  \state{NJ}
  \country{USA}}
\email{pmittal@princeton.edu}

\renewcommand{\shortauthors}{Amos, Maio, and Mittal}

\begin{abstract}

The United Nations Consumer Protection Guidelines lists ``access ... to adequate information ... to make informed choices'' as a core consumer protection right. However, problematic online reviews and imperfections in algorithms that detect those reviews pose obstacles to the fulfillment of this right. Research on reviews and review platforms often derives insights from a single web crawl, but the decisions those crawls observe may not be static. A platform may feature a review one day and filter it from view the next day. An appreciation for these dynamics is necessary to understand how a platform chooses which reviews consumers encounter and which reviews may be unhelpful or suspicious. We introduce a novel longitudinal angle to the study of reviews. We focus on ``reclassification,'' wherein a platform changes its filtering decision for a review. To that end, we perform repeated web crawls of Yelp to create three longitudinal datasets. These datasets highlight the platform's dynamic treatment of reviews. We compile over 12.5M reviews---more than 2M unique---across over 10k businesses. Our datasets are available for researchers to use.

Our longitudinal approach gives us a unique perspective on Yelp's classifier and allows us to explore reclassification. We find that reviews routinely move between Yelp's two main classifier classes (``Recommended'' and ``Not Recommended'') -- up to 8\% over eight years -- raising concerns about prior works' use of Yelp's classes as ground truth. These changes have impacts on small scales; for example, a business going from a 3.5 to 4.5 star rating despite no new reviews. Some reviews move multiple times: we observed up to five reclassifications in eleven months. Our data suggests demographic disparities in reclassifications, with more changes in lower density and low-middle income areas. Because our web crawls coincided with the COVID-19 pandemic, our data also allowed limited exploration of the impact of mask policies and discussions on reviews.

\end{abstract}


\maketitle

\newcommand{\MukherjeeRestaurantCt}{130}
\newcommand{\MukherjeeHotelCt}{85}
\newcommand{\OurRestaurantCt}{129}
\newcommand{\OurHotelCt}{79}
\newcommand{\MukherjeeReviewMatchPerc}{91.8\%}

\newcommand{\RecToFil}{3.9\%}
\newcommand{\FilToRec}{41.3\%}

\newcommand{\OriginalFilteredPercentage}{12.5\%}
\newcommand{\NewFilteredPercentage}{11.6\%}

\section{Introduction} \label{sec:introduction}

Online reviews are an important source of consumer information, play an important role in consumer protection, and have a substantial impact on businesses' economic outcomes~\cite{luca2016reviews,anderson2012impact,un2003conpro}. This creates incentives for various parties to engage in problematic reviewing practices~\cite{streitfeld2012buy,miller19plastic}. These problematic reviews can encompass a large spectrum of behaviors, such as the creation of new accounts for posting fake reviews, hijacking legitimate accounts, compensating real users for posting favorable reviews, incentivized reviews, reviews written by friends or competitors, hiding of negative reviews, requesting negative reviews be handled confidentially, or reviews that are not relevant to the product or service. This is subject of ongoing interest to regulators~\cite{jindal2008opinion,yelpwhyrec,ftc21notice,ftc21fashion,yeung2021Bad}. To address the challenges posed by problematic reviews, review platforms have created classification systems to present users with the best reviews to make informed decisions. A wealth of prior work has taken a variety of approaches to understanding online reviews and the underlying classification systems. This work ranges from studying the factors motivating both honest and problematic reviews to exploring how to detect and generate opinion spam or fake reviews~\cite{jindal2008opinion,yoo2008motivates,baginski2014exploring}. Most of this work examines reviews and review classification systems from a single time-point; however, this perspective is incomplete---these classification decisions are not static.

\textbf{Contributions.} In this work, we move from a view of reviews at rest to a view of reviews in motion: how do reviews move between classes? We explore the dynamic classification of reviews through the collection and analysis of three novel, longitudinal datasets, focused on Yelp. We sought to observe the changing nature of review classification, and thus we collected datasets in which we observed the reviews for a fixed set of businesses across multiple time-points. Our datasets total over 12.5 million total reviews and two million unique reviews, with observations over timescales ranging from four months to eight years. The longitudinal aspect of data allows us to observe the movement of reviews between classifications, which we call ``reclassification''. We take advantage of our dataset to study other aspects of online reviews: our carefully chosen cross sections allow us to approach questions of demographic interactions with reviews, and our timing allows us to explore questions around COVID-19's impact on reviews.

Our contributions are as follows:
\begin{enumerate}
    \item The largest longitudinal dataset of reviews, tracing over 2M reviews across over 11k businesses.
    \item The first study of review reclassification and exploration of this yet-unstudied phenomenon. We find reclassification rates between 0.54\% over four months to 8.69\% over 8 years. Furthermore,  newer reviews are more affected, and that classification appears to happen mostly at an author level.
    \item An exploration into the impacts of income and density, showing disparities in review frequency and reclassification by both income and density.
    \item An investigation of the impacts of mask discussion and rules, showing that the classifier erases rating differences for businesses requiring masks.
\end{enumerate}

\textbf{Implications.} Our results demonstrate that reviews are routinely reclassified, occasionally multiple times. Understanding these reclassifications helps to highlight consumer issues around review classifiers and the challenging nature of building robust classifiers. The magnitude of reclassification also calls into question the validity of the studies that depend on Yelp's classification labels as ground truth~\cite{rayana2015collective,kc2016temporal,mukherjee2013yelp,zhu2021ifspard,shehnepoor2017netspam,yao2017automated}. Multiple-reclassifications (reviews with 2+ reclassifications) suggest that reclassifications do not always move towards ground truth, especially considering the 1,233 multiple-reclassifications we identified. Balancing the issues of fairness to legitimate reviewers with the need to remove problematic reviews is a challenging problem, and our work sheds light on the challenges faced.

\section{Related work} \label{sec:related_work}

Extensive academic literature from multiple disciplines studies online reviews and fake reviews. We highlight four primary areas of prior work: longitudinal study of reviews, demographics and reviews, problematic reviews, and incentives for reviewing.

\textit{Longitudinal study.} Some researchers have performed longitudinal analyses on reviews with a single snapshot, for example by using review dates~\cite{bakhshi2014demographics,ye2016temporal,wang2017temporal}.  Other studies have linked datasets together to gain a better vantage on the review landscape, but still rely on a single snapshot for each data point. For example, \citet{nilizadeh2019think} used reviews from multiple different review platforms along with change point analysis to find fraudlent reviews. To the best of our knowledge, no other studies focus on review reclassification. Yelp acknowledges that reviews are classified by an automated system and sometimes reclassified, but Yelp does not disclose the frequency with which this happens \cite{yelpwhyrec,yelpwhychange}.

\textit{Demographics and reviews.} Ensuring equal access to and treatment by technology is an important equity issue. \citet{baginski2014exploring} explored the hypothesis that, within Franklin County, OH, low income areas have fewer reviews. Instead, they found that there were strong concentrations of reviews, and suggested that technological adoption may play a role. Van Velthoven et al.~\cite{van2018cross} support this hypothesis in their work exploring reviews and ratings in a medical setting; they also do not find a strong link between income or urban/suburban living and the frequency of review authorship. In contrast, \citet{bakhshi2014demographics} find that, among restaurant reviews, local population density has a small but statistically significant effect on review count, but not on rating.  However, \citet{sutherland2020topic} show, using topic modelling, that in hotel ratings rural and metropolitan settings and decor are important discussion points for consumers. Our work helps improve the perspective on how demographics shape reviews.

\textit{Problematic reviews.} Problematic reviews have served as a persistent challenge in the review landscape, largely in the context of detecting fake reviews. \citet{ott2012estimating} estimated that fake reviews occurred at a rate of 2-6\% across six platforms, but other estimates of fake reviews reach 50\%-70\%~\cite{dwoskin2018merchants,elliott2018trust}. Yelp reports filtering about one quarter of its reviews~\cite{yelp2010recommend}. A wealth of research focuses on the problem of detecting these fake reviews~\cite{jindal2008opinion,martens2019towards,ye2016temporal,shehnepoor2017netspam,kumar2018rev2,harris2012detecting,mukherjee2013yelp}. Some works have taken an adversarial approach, generating fake reviews rather than detecting~\cite{adelani2020generating,juuti2018stay,yao2017automated}.

The filtering of problematic reviews introduces new challenges. \citet{eslami2019user} show that some users find the automated filtering system on Yelp to be frustrating and discouraging. However, many users also view the system to be an essential protection against problematic reviews. Many users disliked the opacity of filtering decisions.

Some prior work has attempted to audit existing problematic review classifiers. For example \citet{kamerer2014understanding} attempted to identify both review-based features and reviewer based features that predicted a Not Recommended classification on Yelp. Based on the features they examined, they find that reviewer-based features are more predictive of classification. Similarly, \citet{mukherjee2013yelp} found that, while there are linguistic differences between Recommended and Not Recommended reviews, the reviewer-based features were more predictive of Yelp's classification.

One challenge in analyzing fake reviews is the absence of ground-truth data. Fake reviews may be designed to fool even humans~\cite{ott2011finding}, and only the author of a review may know its authenticity with certainty. \citet{wang2016real} obtained ground-truth data by using leaked data from fake reviewers. \citet{martens2019towards} posed as customers to fake review providers to identify other fake reviews. \citet{ott2011finding} had study participants create fake reviews. Other studies rely on suspected---but unconfirmed---fake reviews. ~\citet{jindal2008opinion} used ``obviously'' fake reviews (e.g., duplicates) on Amazon to train a model to find other fake reviews. \citet{mukherjee2013yelp} and \citet{rayana2015collective} rely on Yelp's classifications for analysis.

Our work most closely resembles the work of \citet{kamerer2014understanding} and \citet{mukherjee2013yelp}, in that we study Yelp's deployed classifier, particularly focusing on the dynamic nature of decisions made.

\textbf{Incentives.} A number of economics studies have tried to understand the incentives behind posting both legitimate and problematic reviews. \citet{yoo2008motivates} showed that most consumers post reviews for themselves, to help the company, and to protect other consumers, while a smaller portion do so as retribution for poor service. \citet{luca2016fake} studied the economic incentives for problematic review posts, concluding that chain restaurants, restaurants with stable ratings, and restaurants with not much competition are less likely to have fake reviews posted on their pages. This literature is crucial for interpretation of the results of our study.

\section{Data collection} \label{sec:dataset}

We collected and constructed three longitudinal datasets to study reviews on Yelp. We present background on Yelp in Section \ref{subsec:background}, we describe the difference between our three datasets in Section \ref{subsec:target_set}, we describe our crawling process in Section \ref{subsec:crawling}, we describe our data organization steps in Section \ref{subsec:organization}, and how to access our data in Section \ref{subsec:availability}.

\subsection{Background} \label{subsec:background}

Yelp breaks reviews into two primary categories, assigned by a software classifer. The categories are ``Recommended'' and ``Not Recommended''. Yelp lists four reasons for classifying a review as ``Not Recommended'': conflicts of interest, solicited reviews, reliability, and usefulness. Not Recommended reviews do not affect metrics and are displayed less prominently than Recommended reviews~\cite{yelpwhyrec,yelprecommendationsoftware,yelpstarrating}. In order to study the classifier, it is important that we collect both Recommended and Not Recommended reviews. Yelp also presents a third class for reviews: ``Removed for Violating our Terms of Service'', which we do not use in our analysis. While Yelp has published an official review dataset for academic purposes~\cite{yelpacademicdataset}, this dataset is not up-to-date and does not include Not Recommended reviews.

We chose to study Yelp because prior work had established reference datasets we could compare against; we chose to use \citet{mukherjee2013yelp}'s dataset of Yelp reviews, collected in 2012, which contains both Recommended and Not Recommended reviews from around 200 restaurants and hotels in Chicago. Furthermore, unlike many other platforms, Yelp allows access to reviews that it does not recommend.

\subsection{Target set selection} \label{subsec:target_set}
Selecting the target set, the set of zipcodes or businesses to study, required careful selection of sample.

\textbf{Coarse crawl (EYG).} 
The first dataset was a single crawl in which we recrawled the same businesses focused on by \citet{mukherjee2013yelp} Thus, our sample was fixed by the original crawl. Specifically, the Mukherjee et al.~crawl collected all businesses from a target set, then they collected all reviews from the accounts which posted on the targeted businesses, finally they collected metadata for the businesses from those posts. We re-crawled Mukerjee et al.'s target set of businesses, since those are the businesses for which we have the most complete data. We call this the ``eight year gap (EYG)'' crawl because Mukherjee et al.~performed their crawl in 2012 and we performed ours in 2020. By comparing our crawl against Mukherjee et al.'s crawl, we are able to observe reclassifications in the reviews.

\textbf{Fine crawl (CHI).} While our coarse dataset shows changes over a long time scale, it does not reveal how frequently reclassifications occur. To address this, we built a second, finer grained dataset by repeatedly collecting reviews 8 times over 11 months. We chose to use the same zipcodes so that there would be some intersection with the EYG crawl businesses, ensuring continued crawling of some EYG crawl businesses which helps contextualize those businesses. Because the zipcodes are within a single metropolitan area, the Chicago area, we call this the ``Chicago (CHI)'' crawl. This more comprehensive but localized coverage of reviews allows for reviewers to be observed posting multiple reviews.

\textbf{Population Density and Income (UDIS / UDS \& UIS).}
Our fine grained dataset gives insight into a local review ecosystem, but it is possible that the sample chosen is not a representative sample. To address this, we collected a third dataset to obtain a broader range of reviews across the US. To allow for the study of reviews in a diverse set of regions, we stratified regions along two axes: one of density, one of income. We collected density and income information from the US Census~\cite{acs2019householdincome}, and used ZipCode Tabulated Areas (ZCTAs) as a proxy for zipcode. ZCTAs approximate USPS ZipCodes, and typically, but not universally, match them~\cite{Census2020zctas}.

For the density stratified crawl, we stratified zipcodes into 5 strata, dividing the strata evenly by population, using US Census data for population estimates~\cite{acs2019populationtotal}. We uniformly sampled zipcodes from each strata until we had sampled at least 500 businesses from that strata, using the Yelp Fusion API to help us determine how many businesses were in each zipcode. We then collected 4 monthly crawls of each dataset. We repeated the same process for the income stratified crawl.

The strata for the income crawls are: \$0--\$55k, \$55k--\$68k, \$68k--\$82k, \$82k--\$105k, \$105k--\$250k. The strata for the density crawls are: 0--67 ppl/$\text{km}^2$, 67--302 ppl$/\text{km}^2$, 302--881 ppl$/\text{km}^2$, 881--1,873 ppl$/\text{km}^2$, 1,873--57,541 ppl$/\text{km}^2$.

Since the union of the income and density data is also a useful dataset as broader sample than the CHI dataset, we present some analyses with individual datasets, ``US Density Stratified (UDS)'' and ``US Income Statified (UIS)'', and some with the combined dataset, ``US Density and Income Stratified (UDIS)''.

\subsection{Crawling}\label{subsec:crawling}

\begin{figure}[b!]
    \centering
    \includegraphics[width=0.9\columnwidth]{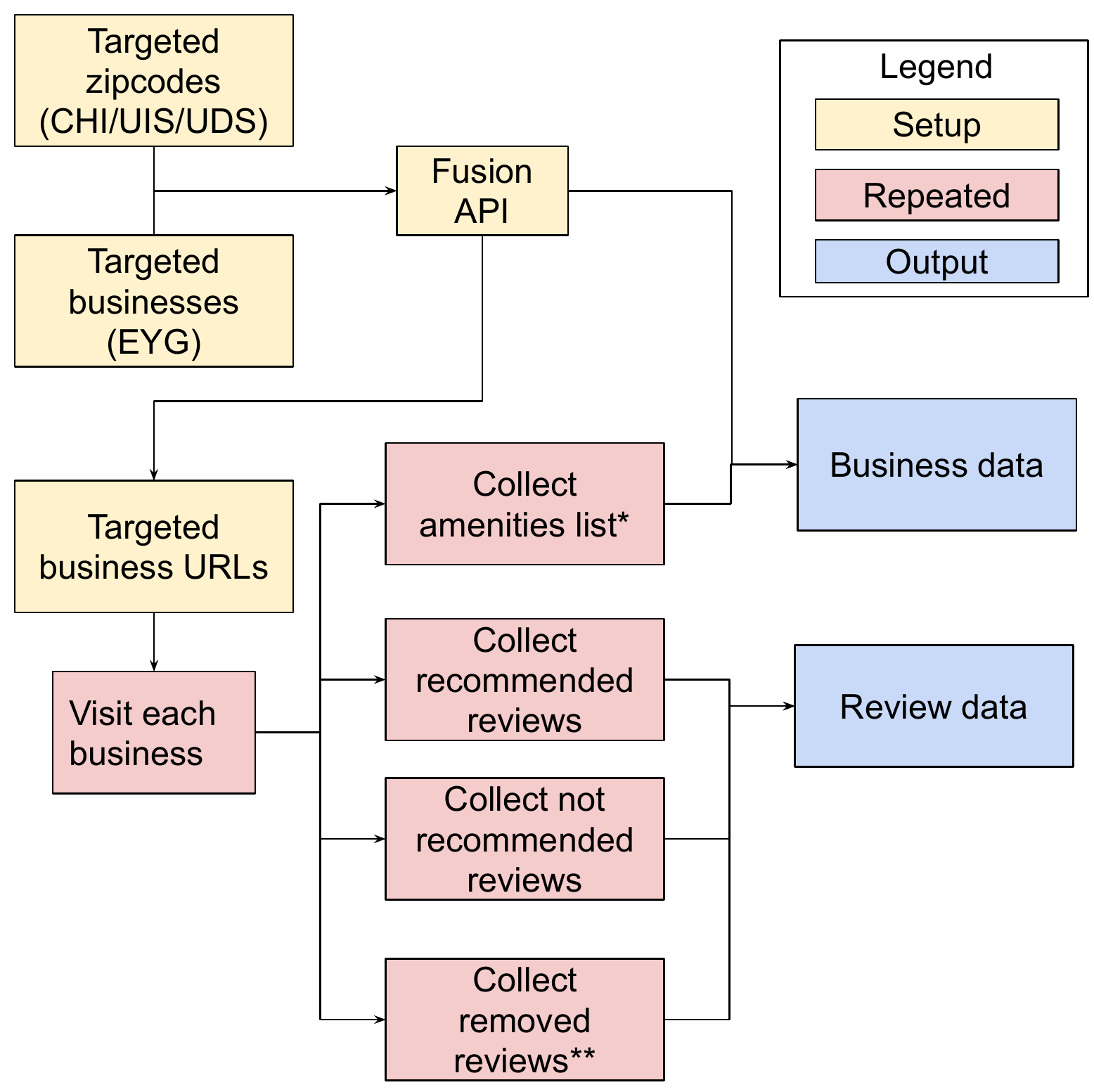}
    \caption{The data collection process. Yellow indicates setup steps that are completed once. Red indicates steps that are completed for each timepoint. Blue indicates outputs.\\
    * Amenities were only collected for the CHI 8 and UDIS 4.\\
    ** Removed reviews were collected for CHI 7-8 and UDIS 3-4. }
    \label{fig:crawling_diagram}
\end{figure}

\begin{table}[t]
    \centering
    \caption{Data and metadata collected.}
    \label{tab:data_collected}
    \resizebox{\columnwidth}{!}{
    \begin{tabular}{l|l}
		  \toprule
            Field & Description \\
			\midrule
			Reviews\\
			\midrule
			Content & Text of the review\\
			Author ID & Reviewer identifier (differs for Recommended / Not Recommended reviews)\\
			Date & Date of posting\\
			Rating & Review rating \\
			Business ID & Identifier for the business the review was posted to\\
			Author data & Name and other public account information\\
			Recommended & Whether the review is Recommended\\
			\midrule
			Businesses\\
			\midrule
			Business ID & Identifier for the business\\
			Amenities & Listed amenities\\
			\bottomrule
    \end{tabular}
    }
\end{table}

Our crawling occurs in two phases.
First, we have an initial setup phase to collect the set of businesses to crawl. Then we have a crawl phase, where we repeatedly collect reviews. At each crawl timepoint, we visit each of the targeted businesses to collect all reviews on that business. We provide an overview of our crawling process in Figure \ref{fig:crawling_diagram}, and a list of data and metadata collected in Table \ref{tab:data_collected}. 

\textbf{Business data}
To collect our data from Yelp, we first needed to identify the businesses in the target set. For the EYG crawl, we used Yelp's Fusion API~\cite{yelpfusion} to collect business URLs for the business identifiers we had. For the CHI and UDIS crawls, for each targeted zipcode we used Yelp's Fusion API to collect a list of all businesses. In situations where the search exceeded the API's response limit, we divided our query into multiple queries using other search parameters to reduce the size of the response. We took the union of the businesses returned by all queries and excluded any results that did not have an address with a targeted zipcodes. For each experiment, once our targeted business list was determined, it remained static for the duration of the experiment.

\textbf{Technologies.}
We used Pyppeteer~\cite{miyakogi2019May}, a Python port of Puppeteer, to build our webcrawler. We ran our crawler in headless mode to reduce system resource utilization. To mitigate IP bans for crawling, we performed our crawl over a VPN.

\textbf{Crawling procedure.}
The crawling process was as follows: we iterated over each zipcode, then each business (in a non-deterministic order). 
We navigated to the business's page, navigating through the list of Recommended reviews. If we detected any inconsistencies in the page, we retried crawling the business. If we received a block page or exceeded 100 page loads since we changed our VPN connection, we connected to a new VPN server. We then navigated to the Not Recommended reviews, where we repeated the same process to collect Not Recommended and Removed reviews. Yelp added an option to include vaccine and mask requirements in early August, 2021 \cite{yelp2021vaccination}. For crawls beginning after mid-August, 2021, we collected the list of ``amenities'', which includes mask and vaccine requirements. 

\begin{figure}[h!]
    \centering
    \includegraphics[width=0.9\columnwidth]{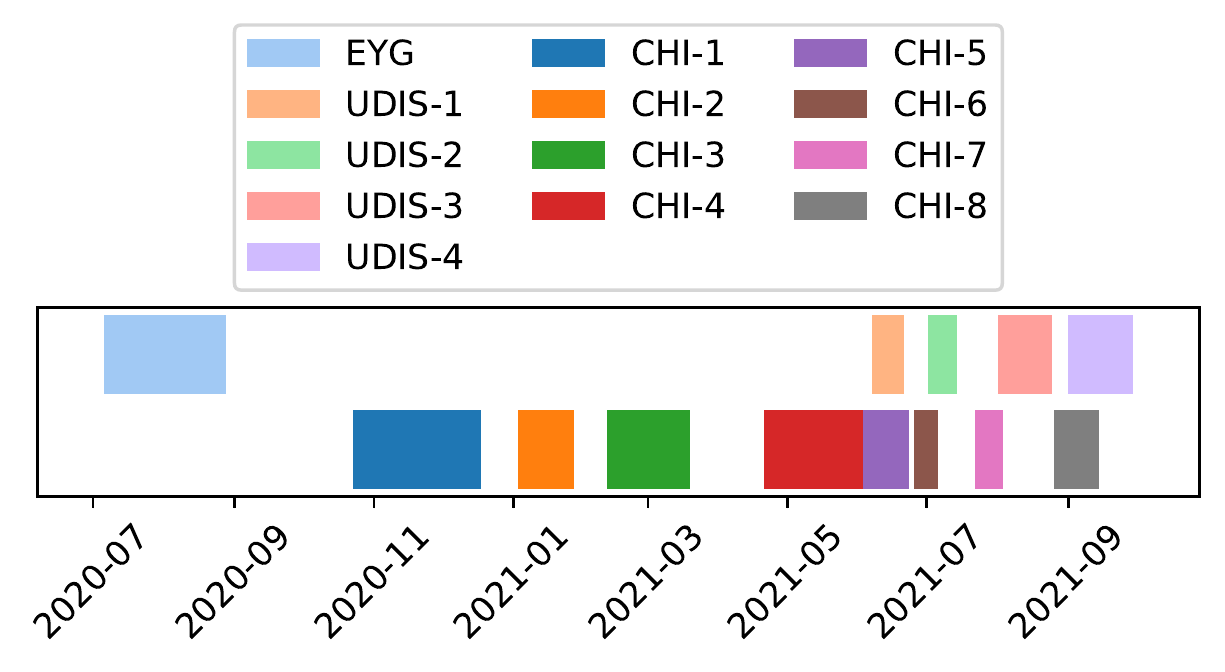}
    \caption{The timeline for each crawl. Each box indicates the first and last operation for each crawl.}
    \label{fig:crawl_timeline}
\end{figure}

The CHI and UDIS crawls were repeated multiple times to allow for a longitudinal perspective. We refer to crawl time-point by the dataset name and 1-indexed count (e.g. CHI-3 is the third crawl of the CHI dataset). We show the timeline of the crawls in Figure \ref{fig:crawl_timeline}.

\textbf{Quality checks.}
Prior work has shown that web crawls using automation tools and headless browsers are easily detectable, and a website could choose to alter content delivered to automated clients \cite{jueckstock2021towards}. In light of this, we performed two checks to ensure the quality of our data. First, we performed an automated check to see review attrition and introduction. Let $R_A$ be the set of reviews for crawl A. For each pair of crawls $\left(A,B\right)$, we checked $\frac{\left|R_A \setminus R_B\right|}{\left|R_A\right|}$ and $\frac{\left|R_B \setminus R_A\right|}{\left|R_B\right|}$. These values never exceed 4.5\% for any pair of adjacent crawls, nor 11\% for any pair of crawls, for either UDIS or CHI. Second, we did a manual check to ensure we collected reviews as they appear for a real user. We randomly selected 50 businesses and a random review position in these business. We manually retrieved the review at that position. Of these reviews, 49 were in our dataset, and 1 was posted after our collection ended.

\textbf{Ethics}
We identify two sources of ethical concerns with our study: the first is the privacy of the user data we have collected, and the second is the impact of our research on Yelp's servers. While all data collected is, or at one point was, publicly available, the review authors did not agree to have their data included in the study. In particular, we treat fields like author name, author location, and review text as sensitive. Therefore, we will require researchers requesting sensitive data to provide an adequate justification for access. To minimize the impact of our research on Yelp's servers, we limited the number of simultaneous crawling threads as much as possible, never exceeding six. We throttled our crawlers to reduce the impact, and we built our crawlers to minimize the pages scraped.

\subsection{Post-processing and organization} \label{subsec:organization}

We took some additional steps to clean up and organize our data.

\textbf{Deduplication.}
Our data has some duplicates. It is possible that some of these are real; for example, if the author accidentally submitted the review twice. However, it is also possible that because our crawls were not instantaneous, review order sometimes shifted during crawling, occasionally leading to double collection of the same review. In either case, such reviews may affect the accuracy of the analysis, and thus we removed these reviews.
To remove duplicate reviews, we removed reviews where all fields (e.g. text, author, date) are identical, retaining one copy.

In our CHI-3 crawl,
approximately 85,000 reviews appeared under both Recommended and Not Recommended, and appeared under Recommended for the adjacent crawls (CHI-2/CHI-4). This coincides with a major update to the Yelp recommendation software~\cite{yelp2021updates}. Because this event boosts the number of double reclassifications approximately 80-fold if we treat these reviews as Not Recommended, we keep the Recommended version.

\textbf{Matching reviews.}
We do not have a unique identifier for reviews, so we rely on heuristics to identify instances of the same review across crawls. To determine if two reviews match, we find all reviews with the same text. If two such reviews appear in the same crawl, we discard all reviews with that text, because we cannot disambiguate them (0.04\% of reviews for CHI and 0.04\% for UDIS). Otherwise, we assume the reviews with that text are the same review.

\textbf{Determining authorship.}
Unlike prior work~\cite{mukherjee2013yelp,rayana2015collective}, we were unable to find a universal identifier for authors. Instead, we found two sets of author identifiers: one for Recommended reviews, one for Not Recommended and Removed reviews. This may be due to site design changes on Yelp. We considered matching authors based on metadata but observed too many false positives to consider this approach reliable. However, for authors with at least one reclassified review, the combination of both identifiers serves as a universal identifier. Thus we focus our investigation of authorship on authors with at least one reclassified review.

\textbf{Composition.} After completing the above cleanup and organization steps, we can examine the composition of the datasets. Table \ref{tab:composition} outlines the scale of the datasets after taking the above steps. CHI is the largest dataset in number of timepoints, number of reviews, and number of unique reviews, while EYG has the longest timespan.

\begin{table*}[]
    \centering
    \caption{Composition of the datasets. ``\# reviews'' is the number of reviews collected; each review counts each time it is observed. ``\# unique reviews'' is the number of unique review texts. ``\# businesses'' is the number of businesses for which we observed any reviews. The ``\# authors'' range lower bound assumes all unmatched authors of Not Recommended reviews have a Recommended review in the dataset; the upper bound assumes they do not.
    ``\% Recommended'' is averaged across all time-points. EYG data includes reviews from \citet{mukherjee2013yelp}'s crawl.}
    \label{tab:composition}
        \begin{tabular}{lc|c|c|c}
             & EYG & CHI & UDS & UIS \\
    Timespan & 8 years & 11 months & 4 months & 4 months \\
    \# Reviews & 263,308 & 10,485,007 & 1,409,059 & 1,145,995 \\
    \# Unique reviews & 196,383 & 1,395,870 & 358,184 & 292,107\\
    \# Businesses & 201 & 5,773 & 2,829 & 2,843\\
    \# Time-points & 2 & 8 & 4 & 4 \\
    \# Authors (range)  & 100,713 - 119,037 & 404,706 - 520,195 & 212,348 - 259,862 & 180,994 - 221,591\\
    \% Reclassified & 8.69\% & 0.87\% & 0.54\% &  0.61\%\\
    \% Recommended & 88.19\% & 88.90\% & 85.69\% & 85.22\% \\
        \end{tabular}
\end{table*}

\subsection{Availability} \label{subsec:availability}
Our crawling and analysis software is available at \url{https://sites.google.com/princeton.edu/longitudinal-review-data/}. Our dataset is available for researchers to access, with the text of reviews and authors' information replaced by a unique identifier. If the text of reviews or author data is needed, a special request can be made for that information.

\section{Results} \label{sec:results}

In this section, we explore three key questions:

\textbf{How extensive are review reclassifications on Yelp?} Yelp presents most reviews as either ``Recommended'' or ``Not Recommended'', and Yelp moves reviews between those categories. However, Yelp does not discuss how frequently this movement occurs. Reclassifications are an indicator of the confidence Yelp has in its classifications, the challenging nature of the problem, and the effort Yelp puts into updating its classifier. Furthermore, reclassifications may frustrate consumers and businesses.

\textbf{How do density and income impact reviews on Yelp?} Disparities in reviews in different regions are an important part of understanding equity on review platforms. For example, it could be possible that certain regions are disproportionately targeted by malicious reviews, or that Yelp's classifier is tuned towards a certain subset of regions.

\textbf{How do mask discussions and requirements impact businesses on Yelp?} With the ongoing COVID-19 pandemic, masks have been a controversial topic~\cite{pascual2021toxicity}; how have mask requirements and mask discussion affected businesses?

For our analysis, we use SciPy~\cite{2020SciPy-NMeth} for statistics, Pandas~\cite{mckinney-proc-scipy-2010} for data processing, and Seaborn~\cite{waskom2020seaborn} for visualizations. We excluded businesses which have no reviews. Unless otherwise stated, we use both Recommended and Not Recommended reviews. All p-values have been corrected for 6 hypotheses using the Holm-Bonferroni multiple hypothesis correction method with a significance level of $\text{p}<0.05$~\cite{seabold2010statsmodels}.

\subsection{Review reclassification} \label{subsec:review_reclassification}

Although Yelp has said that it reclassifies reviews~\cite{yelp2010recommend}, it has not specified the frequency or nature of these changes. Reclassification details could hint at Yelp's approach to classification and details of its classifier, such as the features it considers. The nature of the classification changes could indicate whether Yelp errs towards over- or under-filtering. The frequency of reclassification, illustrated in Table \ref{tab:reclassification}, calls into question the validity of the studies that depend on Yelp's classification labels as ground truth~\cite{rayana2015collective,kc2016temporal,mukherjee2013yelp,zhu2021ifspard,shehnepoor2017netspam,yao2017automated}. As such, it is important to dive into this phenomenon and understand the factors connected to reclassification.

\begin{table}[t]
    \centering
    \caption{Reclassification of reviews between 2012 and 2020. Only includes reviews present in both snapshots.}
    \label{tab:reclassification}
    \begin{tabular}{lcc}
		  \toprule
      & Recommended &  Not Rec. \\
			& (2012) & (2012) \\
			\midrule
      Recommended (2020) & 56,048 & 3,566 \\
      Not Recommended (2020)& 2,249 & 5,059\\
			\bottomrule
    \end{tabular}
\end{table}

\textbf{In the long run, reviews are disproportionately reclassified as Recommended from Not Recommended.}
We first approach reclassification in the long timescale with the EYG dataset. Table~\ref{tab:reclassification} shows how reviews have been reclassified between the two snapshots: most reviews receive the same classification in both snapshots, but a significant number of reviews are classified differently between them ($\chi^2$ test with 1 degree of freedom: $\text{p}<<1\text{e}-5$). We note that more reviews are reclassified as Recommended from Not Recommended than vice versa, which is especially interesting considering that 87.6\% of the reviews were Recommended in the 2012 snapshot. Proportionately, we observe that 3.9\% of the reviews that were Recommended in the 2012 snapshot were Not Recommended in the 2020 snapshot, while 41.3\% of the reviews that were Not Recommended in the 2012 snapshot were Recommended in the 2020 snapshot.

\begin{figure}[t]
    \centering
    \includegraphics[width=0.9\columnwidth]{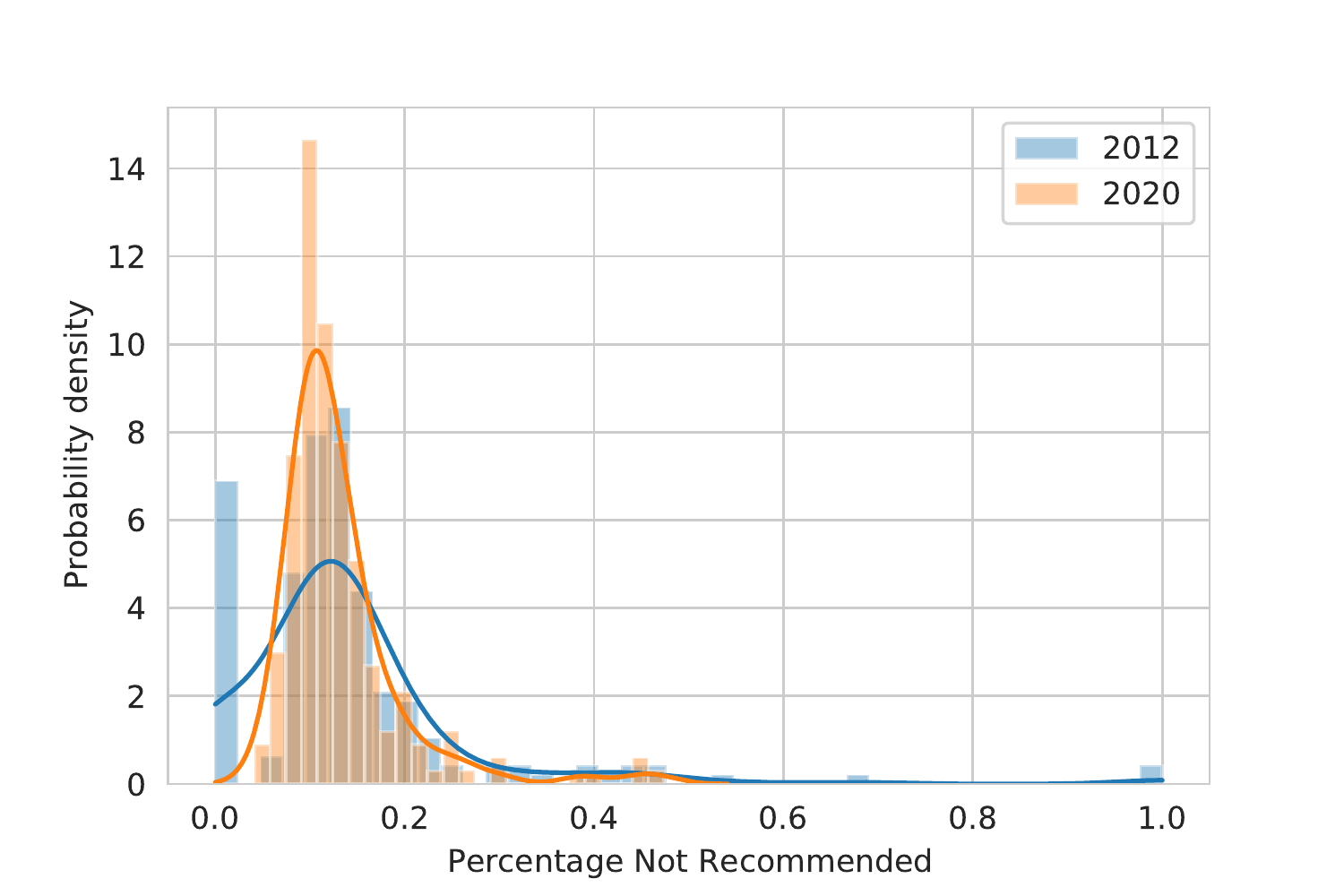}
    \caption{Probability density of Not Recommended review percentage for a business, 2012 and 2020 data. Lines are the kernel density estimates.}
    \label{fig:filtered_density}
\end{figure}

\begin{figure}[t]
    \centering
    \includegraphics[width=0.9\columnwidth]{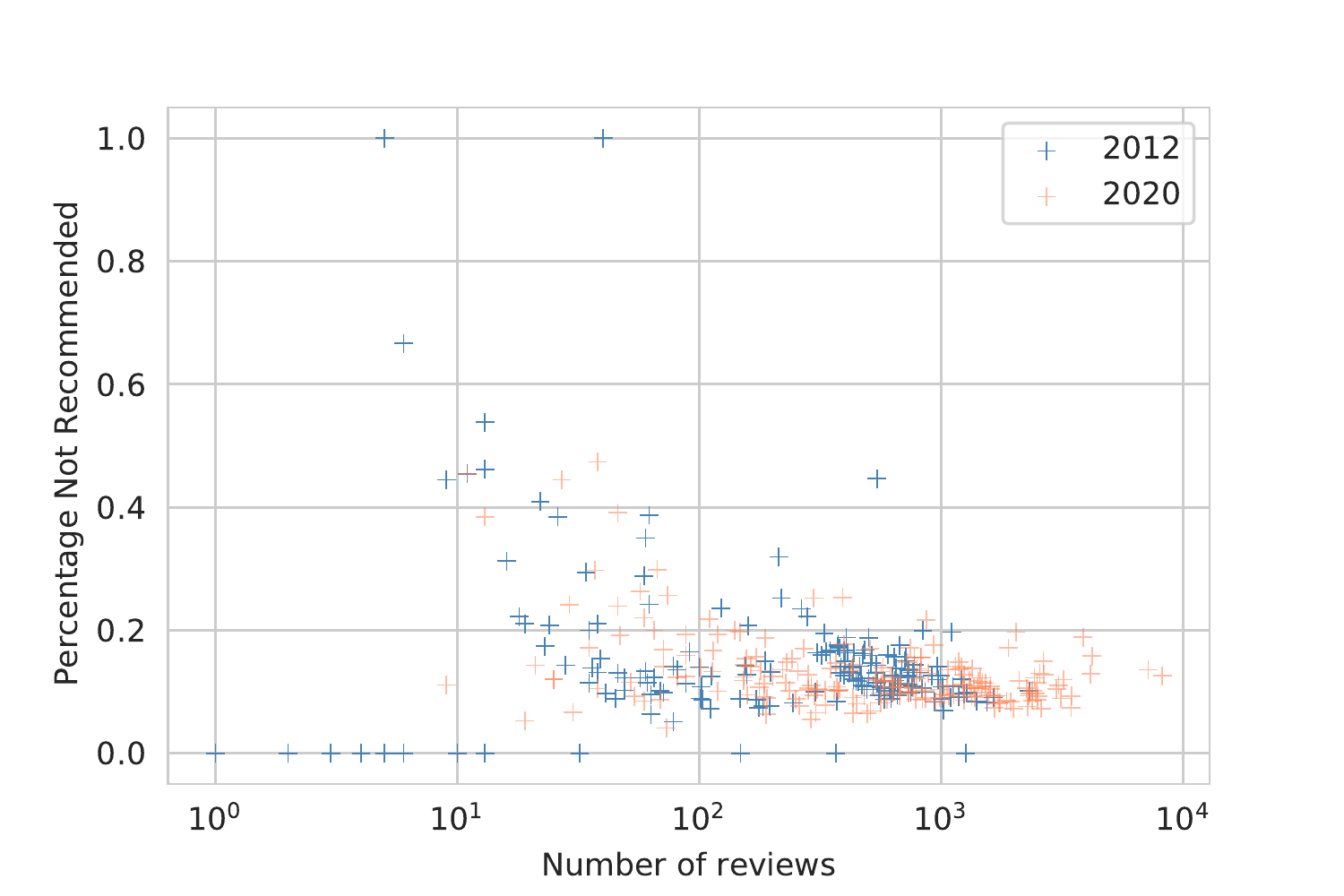}
    \caption{Number of reviews versus the percentage of reviews that were Not Recommended for a business, 2012 and 2020 data.}
    \label{fig:count_vs_perc}
\end{figure}

\textbf{The percentages of Not Recommended reviews per business tend to converge with more reviews.} To understand whether review classification has changed at a business level, we examined the distribution of percentage of Not Recommended reviews per businesses and the connection between number of reviews and percentage Not Recommended per business. Figure~\ref{fig:filtered_density} shows the distribution of percentage Not Recommended by business. The median percentage Not Recommended is similar (0.122 and 0.115), but the distributions are distinct (Kolmogorov–Smirnov test $\text{p}<0.05$). In 2012, a number of business have no Not Recommended reviews. Some of these results may be explained by the business having fewer reviews in 2012 than 2020. Figure~\ref{fig:count_vs_perc} shows the connection between the number of reviews on a business and the percentage Not Recommended. Most businesses converge to around the same percentage Not Recommended with enough reviews. This convergence appears tighter among the 2020 data. We note that there is no significant correlation between the number of reviews and the percentage Recommended for the 2012 data (Spearman's correlation $\rho = 0.13$, $\text{p} = 0.06$), but there is a significant, negative correlation for the 2020 data ($\rho = -0.31$, $\text{p}<1\text{e}-4$)---the more reviews a business has, the smaller the proportion of Not Recommended reviews. The correlation for the 2020 data may be more significant because the businesses are more established.

\begin{table}[t]
    \centering
    \caption{Frequency of reclassification patterns observed in Chicago data over 5 timepoints. ``R'' denotes Recommended, ``N'' denotes Not Recommended.}
    \label{tab:reclassification_patterns}
    \begin{tabular}{llr}
    \# changes & Pattern & Count \\
	 \hline
	 0 & R & 1,235,194\\
	 & N & 148,278\\
	 & \textit{Total} & 1,383,472 \\
	 \hline
	 1& R $\to$ N & 4,953\\
	 & N $\to$ R & 5,573\\
	 & \textit{Total} & 10,526\\
	 \hline
	 2& R $\to$ N $\to$ R & 706\\
	 & N $\to$ R $\to$ N & 373\\
	 & \textit{Total} & 1079\\
	 \hline
	 3+ & R $\to$ N $\to$ R $\to$ N & 60\\
	 & N $\to$ R $\to$ N $\to$ R & 75\\
	 & R $\to$ N $\to$ R $\to$ N $\to$ R & 14\\
	 & N $\to$ R $\to$ N $\to$ R $\to$ N & 15\\
	 & N $\to$ R $\to$ N $\to$ R $\to$ N $\to$ R & 2\\
	 & \textit{Total} & 157\\
	 \hline
    \end{tabular}
\end{table}

\textbf{Many reviews are reclassified, a few are reclassified frequently.} To investigate the frequency and scale of reclassification on shorter timescales we investigate reviews from the CHI dataset. Table \ref{tab:reclassification_patterns} shows how frequently reviews were reclassified. In our study period, around 0.8\% of reviews were reclassified. A small fraction of reviews undergo a substantial number of changes. This is especially interesting in light of the shorter study period and the limited number of time-points---a few reviews changed classes almost every measurement. These reviews may be cases that are particularly hard for Yelp to classify. 

\begin{figure}[t]
    \centering
    \includegraphics[width=0.9\columnwidth]{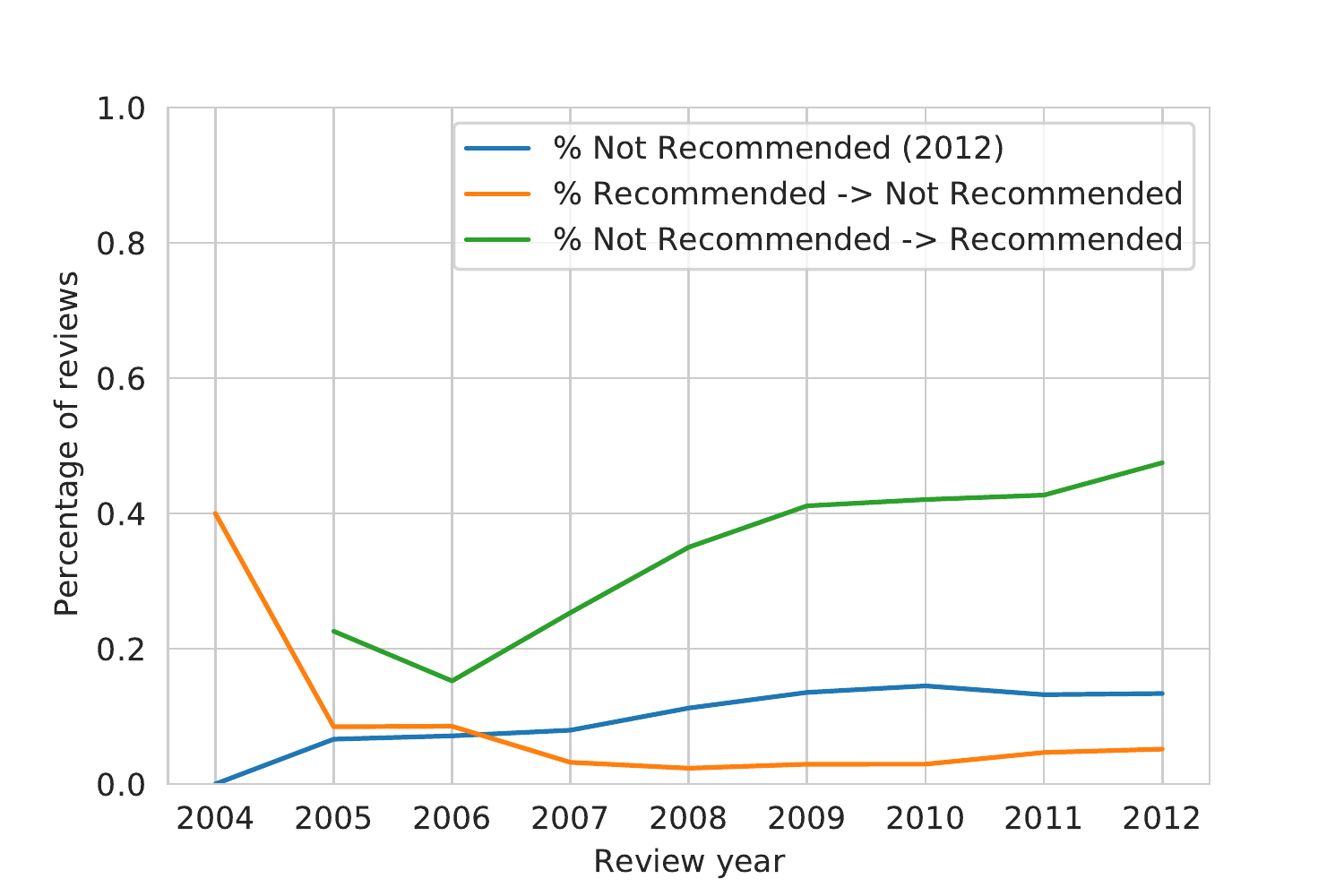}
    \caption{Change in recommendation status by the year reviews posted. The blue line represents reviews Not Recommended in 2012; the orange and green lines represent reviews that were Recommended and Not Recommended, respectively, in 2012 but reclassified in 2020.
    }
    \label{fig:filtered_change}
\end{figure}

\textbf{In the long run, newer Not Recommended reviews are more likely to be reclassified.} To explore the relationship between reclassification and review age, we grouped reviews in EYG by the year posted. Within each group, we measured the percentage of reviews Yelp Recommended in 2012 and the percentage reclassified in 2020 from Recommended and Not Recommended (Figure \ref{fig:filtered_change}). Note that 2004 has just five reviews, and all were Recommended in 2012. The trend of the percentage Not Recommended in both 2012 and 2020 (green) suggests that Yelp was more likely to reclassify newer reviews from Not Recommended to Recommended. Yelp could have had more time to examine older reviews by 2012, or older, less sophisticated fake reviews may have resulted in fewer filtering errors to correct. Alternatively, this could stem from the use of review age or correlated factors (e.g., the number of reviews per author) in classification: Yelp considers whether a reviewer is ``established'' in recommending reviews, which is something Yelp claims to do~\cite{yelpwhyrec,yelprecommendationsoftware}.

\begin{figure}[t]
    \centering
    \includegraphics[width=0.9\columnwidth]{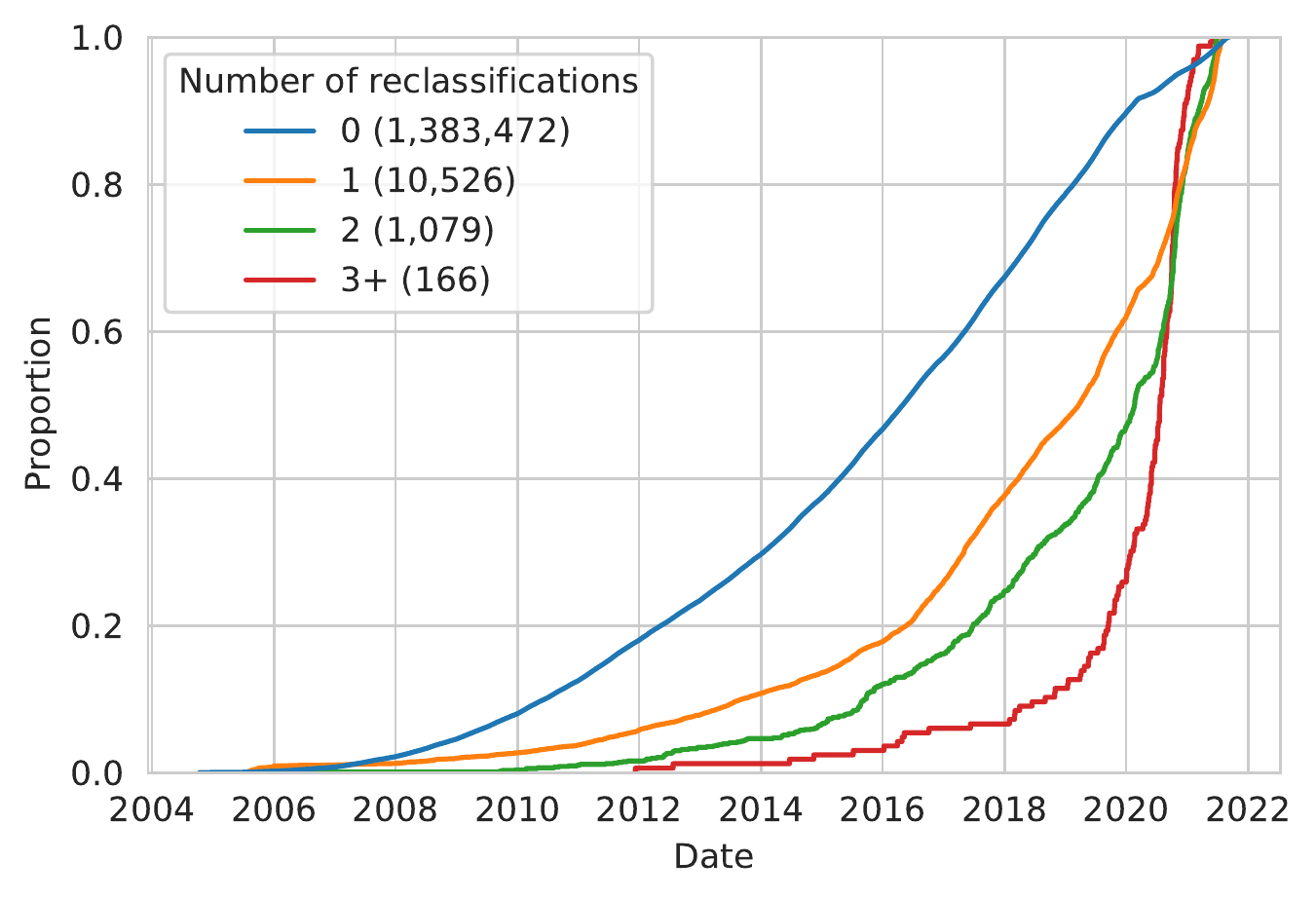}
    \caption{Cumulative percentage of reviews with a given number of reclassifications that were posted by a given date. For example, by the start of 2018 approximately 20\% of reviews with 2 observed reclassifications had been posted.}
    \label{fig:reclassification_by_date_chicago}
\end{figure}

\textbf{Newer reviews are more likely to undergo repeated reclassification, but the chance of reclassification persists over time.} Figure \ref{fig:reclassification_by_date_chicago}, shows reviews which undergo more frequent reclassification tend to be newer. This supports the idea that Yelp increases its confidence in classifying reviews as the reviews age. Perhaps because Yelp has observed more activity from the author's account or the business. We note that we still see reclassifications of reviews dating back to 2005, suggesting that a review's classification is never fully stable. We note a small artifact in the upper-right corner: the 1, 2, and 3+ lines rise above the 0 line. We expect this artifact because the newest reviews cannot have 1, 2, or 3+ reclassifications, since we have not made as many observations of them.

Given these results showing that newer reviews are more likely to undergo reclassification and that the percentage reclassified for the EYG (8.69\% over 8 years or 0.09\% per month) and CHI datasets (0.87\% over 11 months or 0.08\% per month) is similar, it seems likely that Yelp's classifier is either more stable today or that Yelp performed a major overhaul between the EYG collection time-points. It is also possible that the EYG sample was disproportionately subject to reclassification.

\textbf{Review classes follow the author.} In order to determine if reclassifications are performed at the author level, we investigated whether authors who have a review classifcation change are likely to have their other reviews match the new classification. In the 1,175 cases in which an author with multiple reviews had a review reclassified, 924 had all of their reviews match after reclassification. The average percentage of an author's review pairs that matched classification was 95.1\% for Recommended reviews and 94.7\% for Not Recommended reviews. This suggests that classifications follow the author. We also observe some multiple-reclassifications at the author level, indicating misclassifications, which are harmful to legitimate authors whose reviews are hidden. For example, we found an author with 18 reviews temporarily reclassified as Not Recommended. Such double reclassifications indicate an error by the classifier. If the author is a legitimate author, they may be discouraged by this reclassification and choose not to engage in further reviewing. Furthermore, their ability to inform other consumers was greatly diminished during the period where their reviews were classified as Not Recommended.

Reclassification may act as a potential point of frustration or even chilling for legitimate users and businesses, and can lead to questions of fairness -- while the classifier may work well in the average case, the worst case is experienced by real people and thus matters. As two examples of significant swings in rating, we found a business which had its rating go from 3.5 to 4.5 after a reclassification and another from 2.5 to 1, suggesting that either Yelp's initial or updated rating failed to represent legitimate reviewer attitudes.  It also leads to questions about consumer protection in access to information about products -- reclassification shows that consumers are not getting an entirely accurate picture. We did not find that any particular timepoint had significantly more reclassifications.

\textbf{Examples of reclassifications.} To understand the variety of factors preceding a reclassification, next we will examine a few reviews that were reclassified. We focus on reclassifications around account changes -- changes in review count, friend count, and photo count. We observe that reclassifications can occur with no account changes, shortly after account changes, and with a delay after account changes.

Some reclassifications occur despite no observed account changes. For example, we observed two reviews posted in July 2012 and April 2017 with ratings of 1 and 4, friend counts of 0 and 3, and review counts of 3 and 22, respectively. Both were Recommended in CHI 1-4 and Not Recommended in CHI 5-8, despite no account changes.

Other reclassifications occur after an account change. For example, we observed a review posted in October 2020 with a rating of 2 which was Not Recommended in CHI 1-4, but Recommended in CHI 5-8, after the user posted 3 more reviews. This user had approximately 400 friends and 1 review in CHI-1.

Sometimes the changes lag behind account changes. For example, we observed a review from July 2013 with a rating of 5 which was changed to Recommended in CHI 6 after the author posted 2 reviews between CHI 2 and CHI 3. In CHI-1, the user had approximately 180 friends and 9 reviews.

\subsection{Density and income impacts}

We investigated how density and income impact reviews using our UDIS study. Demographic disparities are indicative of issues of fairness in the online review space -- all consumers should have equal access to both write and read reviews relevant to them. While our data is not sufficient to determine the root causes of disparities, we can identify them for further study.

\begin{figure}[t]
    \centering
    \includegraphics[width=0.9\columnwidth]{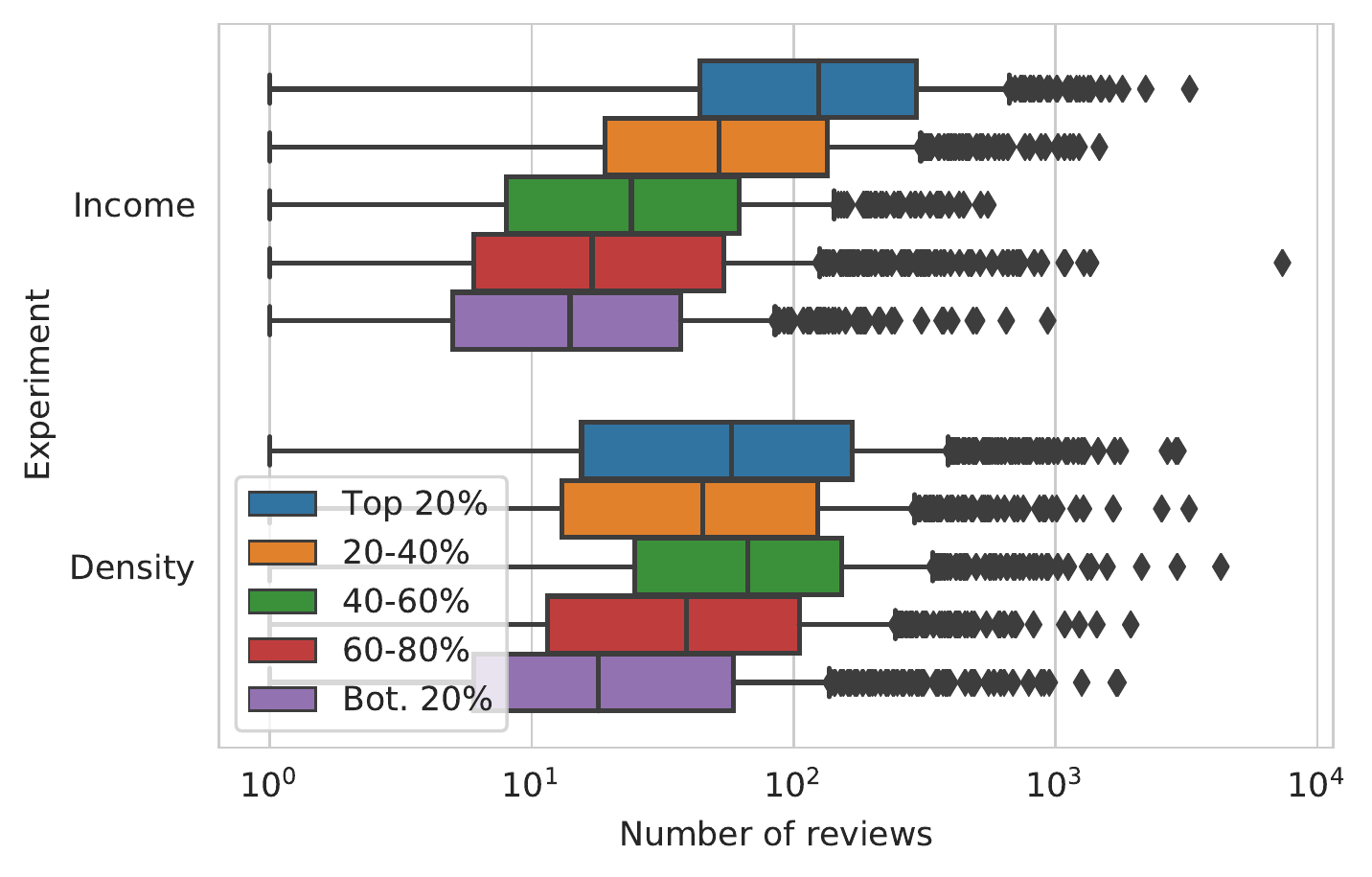}
    \caption{Reviews per business. Each data point is the number of reviews for one business. }
    \label{fig:reviews_per_business_stratified}
\end{figure}

\textbf{Demographic factors correlate with the number of reviews on each business.} We investigated the number of reviews per business in Figure \ref{fig:reviews_per_business_stratified}. Both low income and low density areas have fewer reviews per business than higher income and higher density areas, and the gap is wider for income. This could be because higher income areas typically have businesses with more time on the platform---for the highest income stratum the median oldest review for each business (3,425 days) is 30\% older than for the lowest income stratum (2,638 days). This relationship is not as strong in the density experiment: the middle density stratum has the oldest reviews (3,245 days), slightly higher than the highest stratum (3,066 days) and much higher than the lowest stratum (2,704 days). The disparity in number of reviews means consumers in these areas may have less access to reviews, impacting their ability to make informed decisions.

 \begin{figure}[t]
     \centering
     \includegraphics[width=0.9\columnwidth]{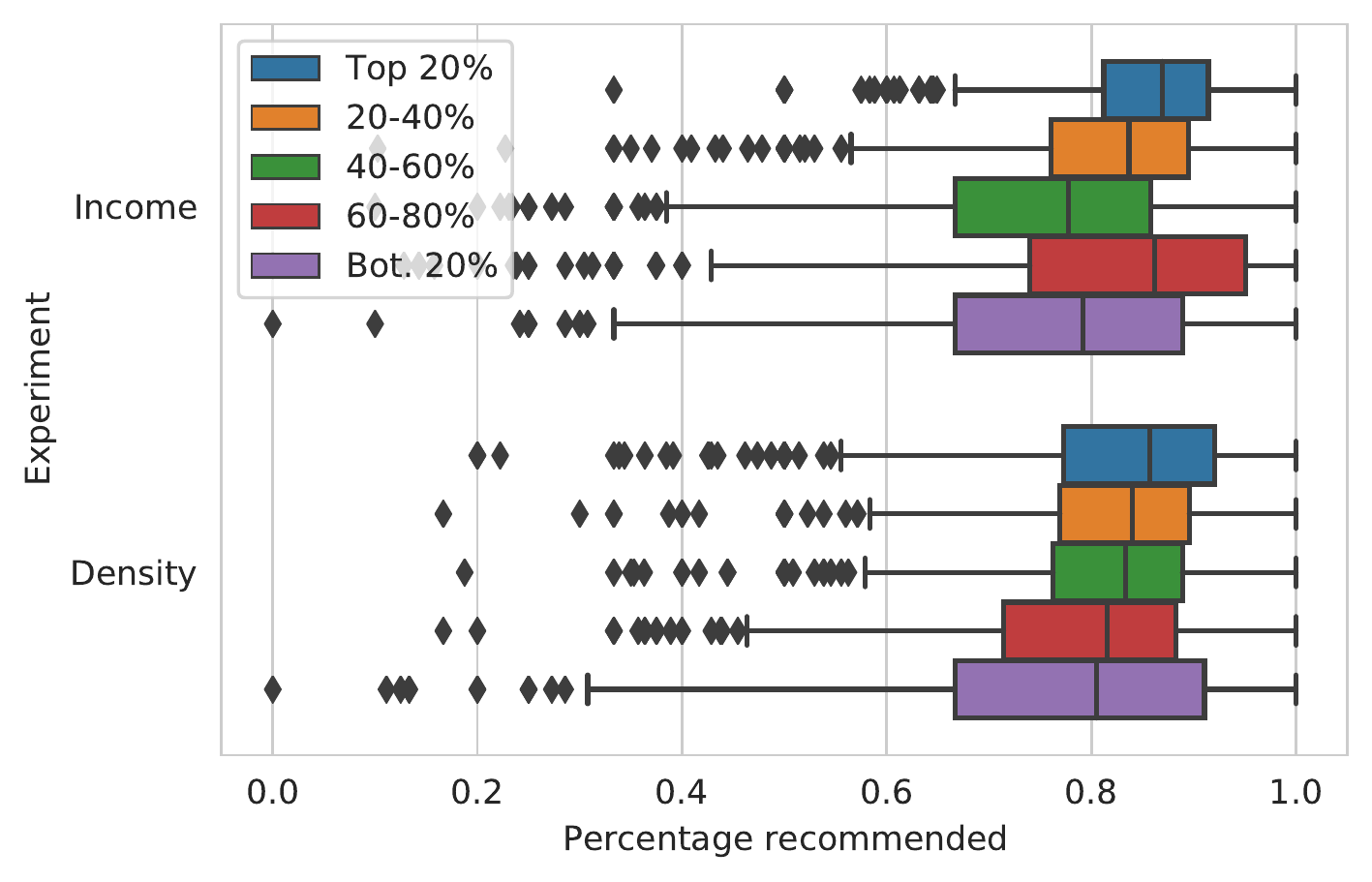}
     \caption{The proportion of reviews Recommended per business. Each data point represents one business.}
     \label{fig:percentage_recommended_per_businesses_extended}
 \end{figure}
 
\textbf{Demographic factors correlate with the percentage of reviews Recommended for each business.} We examined how the percentage of reviews that are Recommended per business varies by income and density in Figure \ref{fig:percentage_recommended_per_businesses_extended}. The range in median percentage Recommended is tighter for density---it ranges from 80\% (Bottom 20\%) to 86\% (Top 20\%)---whereas the range is larger for income, ranging from 78\% (40-60\%) to 87\% (Top 20\%). We observe that higher density and higher income areas generally have a higher percentage of recommended reviews. Possible causes of this range from a lower concentration of problematic reviews to better tailoring of the recommendation algorithms for those areas.

\begin{figure}[t]
    \centering
    \includegraphics[width=0.9\columnwidth]{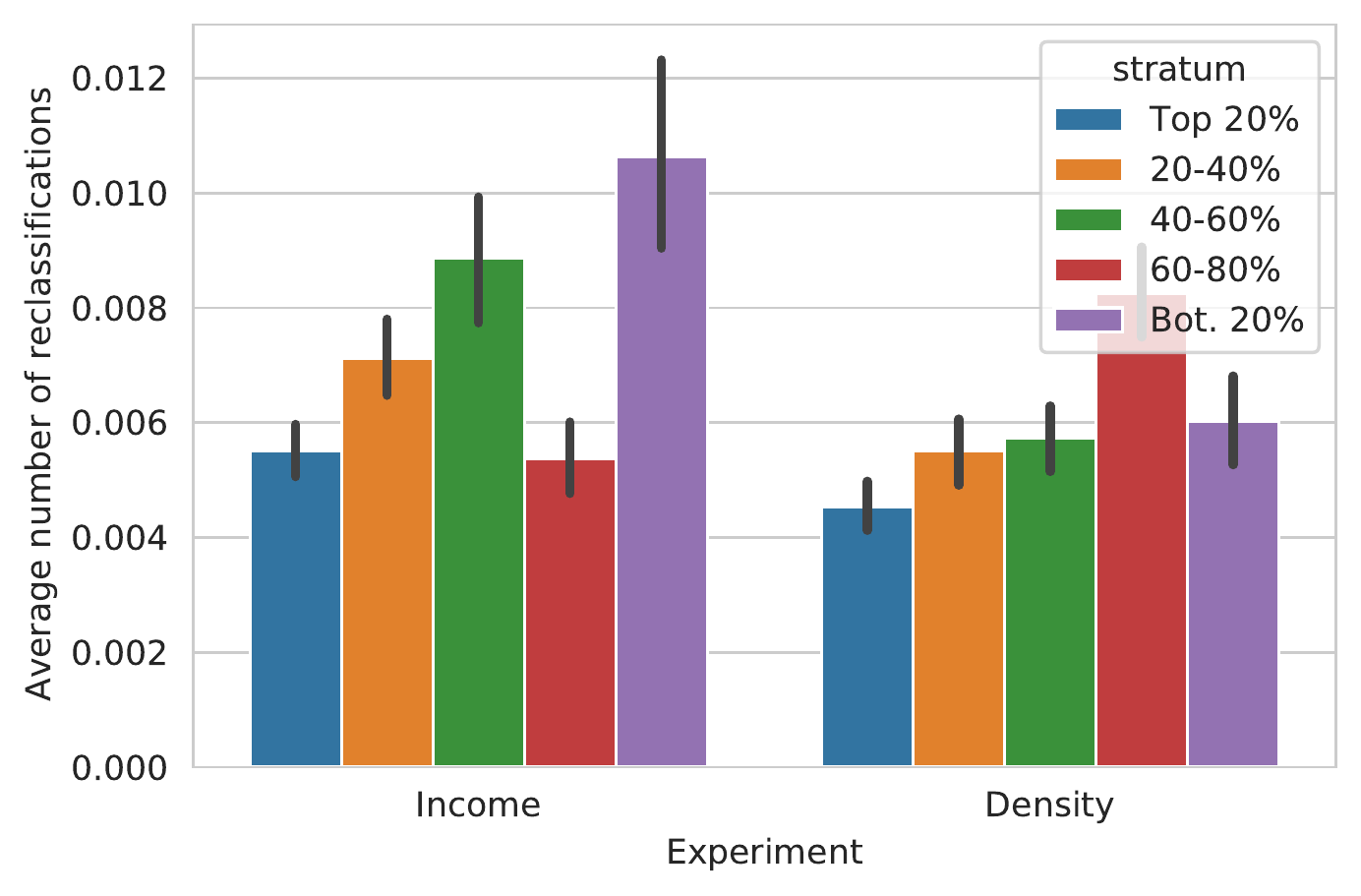}
    \caption{Average number of reclassifications per unique review per stratrum. Black bars indicate the 95\% confidence interval}
    \label{fig:stratified_reclass_swaps_usa}
\end{figure}

\textbf{Demographic factors correlate with the frequency of reclassifications.} In Section \ref{subsec:review_reclassification}, we explored the frequency and factors surrounding reclassification. To test whether Yelp reclassifies reviews for businesses in regions with certain income or density attributes more frequently, we looked at the average number of reclassifications per review in each stratum for both the UIS and UDS crawls (Figure \ref{fig:stratified_reclass_swaps_usa}). Less dense and lower income regions experience more reclassification. The 60-80\% strata are an outlier in both cases---the 60-80\% density stratum experiences significantly more reclassifications while the 60-80\% income stratum experiences significantly less than its neighbors, but on par with the top income stratum, which warrants further research. These disparities invite questions as to why they arise: is there something inherent about these markets that leads to more challenging-to-classify reviews, or is Yelp's classifier not well tuned to them?

\subsection{Masking}

\begin{figure}[t]
    \centering
    \includegraphics[width=0.9\columnwidth]{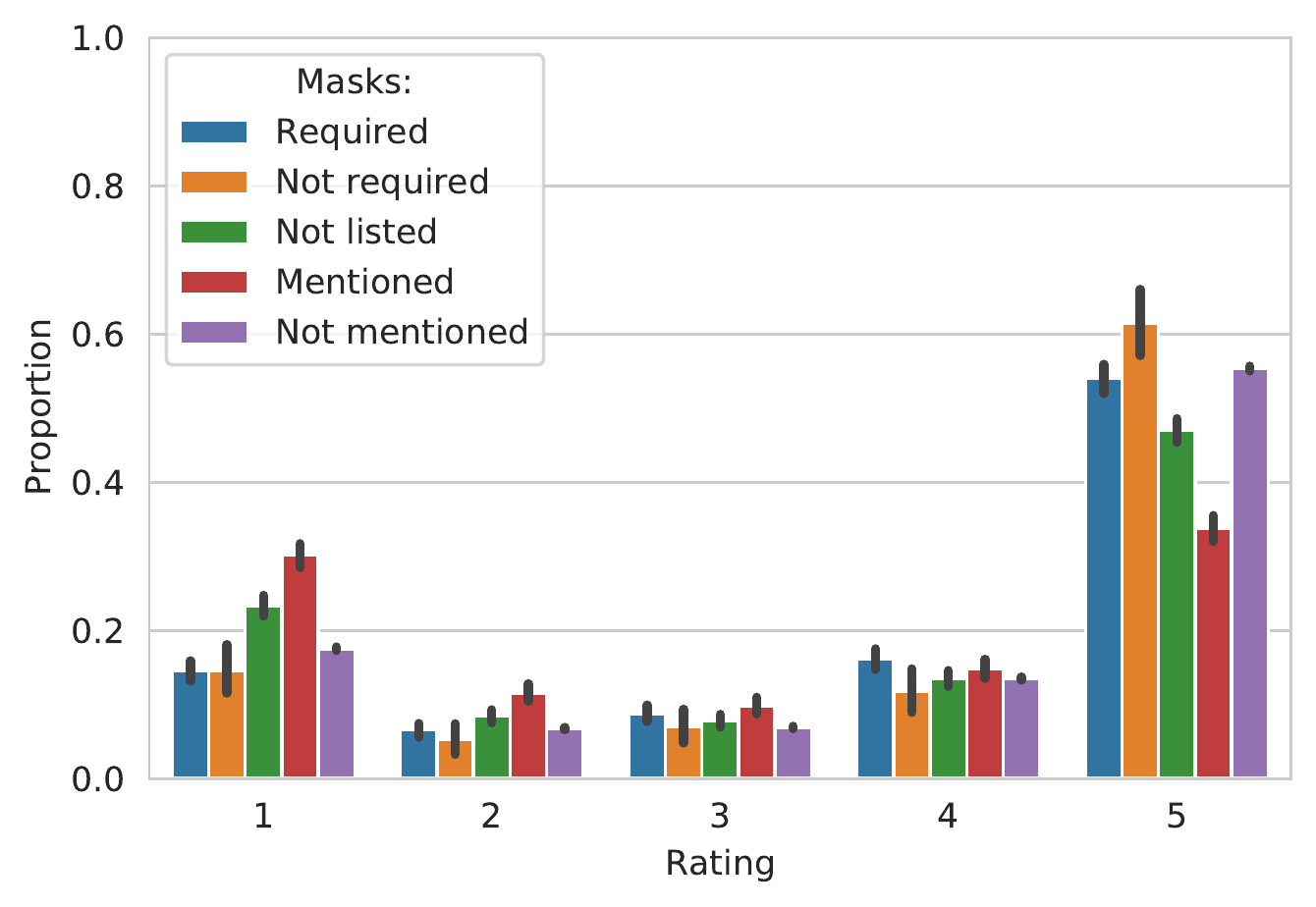}
    \caption{The proportion by rating of reviews on or after August 6, 2021 on businesses that (1) require masks (2) do not require masks (3) do not list a requirement for masks. In addition, the proportion of reviews on or after March 1, 2020 that (4) mention masks (5) do not mention masks. Black bars indicate the 95\% confidence interval.}
    \label{fig:proportion_Masks mentions_masks required_by_rating_usa}
\end{figure}

In response to the COVID-19 pandemic, Yelp added an option for businesses to specify a mask policy in August 2021. Because this coincided with our longitudinal data collection, we studied reviews that mention masks and business with listed mask policies. We used our UDIS-4 crawl data because that crawl started after Yelp added the option. 

We determined whether a review mentioned masks by tokenizing the review and lemmatizing the tokens. If any lemma matched ``mask", we considered that review to mention masks. We find that 90.2\% (2,845) of UDIS reviews mentioning masks occur on or after March 1, 2020. We manually examined a random sample of 20 such reviews from before March 1, 2020: 16 used ``mask'' to describe covering a taste or odor, 1 in a COVID-19 context, and 3 to describe costumes. We examined a random sample of 20 such reviews from on or after March 1, 2020; all 20 of them used ``mask'' in a COVID-19 context. To determine whether a business has a masking policy, we used the business amenities as described in Section \ref{subsec:crawling}. We found 837 businesses requiring masks, 174 not requiring masks, and 4,666 with no listed policy.

\textbf{Mask policies have little correlation with rating, as long as one is present. Discussions of masks correspond to lower ratings.} We show how both customer mask requirements and mask discussion affect rating in Figure \ref{fig:proportion_Masks mentions_masks required_by_rating_usa}. Reviews mentioning masks have a lower rating, and this relationship remains after removing Not Recommended reviews. Having a mask requirement results in a non-significant rating change (means 3.89 and 4.00, Spearman correlation $\rho = -0.04$, $\text{p}=0.06$), and this relationship vanishes after removing Not Recommended reviews (mean 3.92 and 3.90, Spearman correlation $\rho = -0.01$, $\text{p}=0.79$). This suggests Yelp's filter may have a mild effect of protecting restaurants requiring masks. Listing any policy correlates with higher ratings; this could be explained by the overall correlation between higher ratings and more listed amenities: the Spearman's rank correlation between the rating and the number of amenities is $\rho=0.119$ $(\text{p}<<1\text{e}-5)$. While more data is needed to establish statistical significance, our results seem to conflict with the results of \citet{kostromitina2021his}, who found that reviewers generally reviewed more favorably those businesses with better COVID safety protocols. Consumer perceptions of businesses' health and safety protocols are important for policy makers who might wish to rely on the free market rather than statute to prescribe health and safety practices.

\section{Discussion} \label{sec:discussion}
Online reviews are part of an actively evolving landscape with significant economic consequences. In this paper, we have examined this landscape from four different cross-sections: a course-grained eight year view;
a more fine-grained, eleven month view focused on one region; a four month view sampled from the whole US stratified by density; and a second four month view stratified by income. Each of these datasets is available for other researchers to use.

Reviews on Yelp routinely move between classifications, in both directions, occasionally multiple times. Newer reviews are less stable in their classification, but even old reviews are still subject to occasional reclassification, even multiple reclassifications. These reclassifications are often connected to the review author. Density and income are connected to the number of reviews per business, the review classifications, and how frequently reviews are reclassified. We find both discussion of masks and, to a lesser extent, declaring a mask policy, impact the ratings given by reviewers. Our methodology can offer insight into an opaque process for reviewers, consumers, and businesses who might not understand why their review was blocked or why the reviews they see change.

Our results have implications for platforms and policymakers. Our reclassification results demonstrate the uncertainty in the recommendation process. We suggest other platforms consider a greater transparency model with their reviews, similar to Yelp’s model -- that platforms should remove, but still make accessible, reviews believed to be problematic. Furthermore, platforms should be cautious about changing classification until they are confident the review is not problematic. Platforms and regulators should consider carefully any discrepancies by density and income: are there steps that can be taken to address these inequities? Our observations surrounding masking policies suggest public backlash against a business's mask policy decision is negligible, which may be due in part to Yelp's classifier.

\subsection{Limitations} \label{subsec:limitations}

Our study is limited to a single platform and a single country, so it may not be representative of trends on other platforms or other countries. Our study period includes the COVID-19 pandemic, a period of substantial social and economic disruption~\cite{altig2020economic,deb2020economic}. Furthermore, local median income and population density may not completely describe the businesses and reviewers; for example reviewers may travel from another area to the business or the area may be heterogeneous.

We expect that we may have missed reclassifications that occurred between crawl points and our crawler may have missed reviews (e.g. if the reviews reordered mid-crawl due to a new review). This means that some reviews may have been reclassified more frequently than we observed.

We also do not have a complete view of the review space---we have limited information on reviews removed for terms of service violations and no information on reviews removed by their author before our first crawl. Some of these reviews may be reviews of interest---for example, some problematic reviews may be removed entirely instead of made Not Recommended.

\subsection{Future work} \label{subsec:conclusions}
An immediate question that arises is whether the trends we observe hold true on other platforms. In particular, how do these trends translate to platforms with different monetization structures? Platforms like Amazon benefit more directly from sales by the businesses whose reviews they host---will this affect how they approach review classification? Less transparent platforms would likely require more intensive study, for example over longer time periods to observe more review movement. It may be possible to directly study the classifer by injecting researcher generated reviews, but care would need to be taken when navigating the ethical concerns.

Further investigation into the income and density disparities could be impactful for both platforms and regulators seeking to ensure equity and protect consumers who rely on reviews. It is possible that these issues are caused by forces outside of the control of the platform---for example, rural areas have less access to high speed internet \cite{fcc2020broadband}---but it is also possible that there are steps platforms could take to address these issues.

It may be interesting to investigate how reviews and review rates varied pre- and post- COVID-19 vaccine distribution, and our CHI dataset includes data from before the first vaccines were given an emergency use authorization \cite{nature2020moderna}. Mask mentions and business requirements may have an impact on review ratings, so vaccine distribution and business vaccination requirements may have an impact on reviews.

An additional subject of longitudinal study that we did not cover in our study is editing and deletion of reviews and business data. Future work could attempt to answer questions such as: what prompts users to edit or delete reviews? Which accounts are most likely to edit or delete reviews? What changes are made during edits?

We hope that our work brings momentum to the longitudinal study of review platforms. To that end, our data, crawler, and analysis code is available at \url{https://sites.google.com/princeton.edu/longitudinal-review-data/}.

\begin{acks}
 The authors thank Anne Kohlbrenner, Amy Winecoff, Ben Kaiser, and Sayash Kapoor for their help editing the paper. The authors thank Joe Calandrino for his extensive input and assistance with this work.
\end{acks}

\bibliographystyle{ACM-Reference-Format}
\bibliography{ref/ref.bib}

\end{document}